\documentclass{article}

\usepackage{arxiv}

\usepackage[utf8]{inputenc} 
\usepackage[T1]{fontenc}    
\usepackage{hyperref}       
\usepackage{url}            
\usepackage{booktabs}       
\usepackage{amsfonts}       
\usepackage{nicefrac}       
\usepackage{microtype}      
\usepackage{lipsum}
\usepackage{graphicx}
\usepackage{mathtools}
\usepackage{bbm}
\usepackage{tikz}
\usepackage{graphicx}
\usepackage{amsmath}
\usepackage{amsthm}
\newtheorem{proposition}{Proposition}
\usetikzlibrary{positioning,calc}
\graphicspath{ {./images/} }

\title{On the statistical reliability of chaotic simulations at reduced precision}

\author{
Ross Ah-Weng \\
  Centre for Computational Science\\
  University College London\\
   \And
 Peter Coveney \\
  Centre for Computational Science\\
  University College London\\
}

\begin{document}
\maketitle
\begin{abstract}
Reduced-precision arithmetic is increasingly used to accelerate scientific simulation, while AI workloads are driving its growing prominence in hardware design. Yet small arithmetic errors do not necessarily imply accurate statistics. Finite representation limits spatial resolution, while rounding can merge trajectories that should separate in chaotic systems, progressively concentrating probability on a shrinking collection of computer states. This redistribution can distort statistical averages well before periodicity becomes apparent; furthermore, disjoint computational basins can induce an initialisation bias, even when the intended equilibrium statistics are independent of the initial ensemble. We use functional graphs and their transfer operators to connect these pathologies to bounds on statistical bias, the duration of accurate averaging, and the effect of reducing precision. Exhaustive numerical experiments on four chaotic maps, with $1-24$ significand bits and the $24$-bit endpoint matching IEEE FP32, show how initially accurate ensembles drift towards incorrect statistics, with lower precision often bringing this deterioration forward. Stochastic rounding substantially extends some useful averaging windows but leaves others severely biased. Longer computed trajectories can therefore worsen inferred statistics, and a finite average can outperform the "infinite-time" computer average by orders of magnitude. This exposes a risk of statistically biased inference at reduced precision and provides a further motivation for collecting statistics from ensembles of shorter trajectories.
\end{abstract}


\section*{Introduction}

Consider a simulation of the atmosphere or of a molecular ensemble, and lower its floating-point precision. Are the forecasted statistics the same, or even trustworthy? Two developments make this question pressing: AI workloads are shaping the development of accelerator designs around lower-precision computation, while scientific applications seek gains in throughput, memory use and energy efficiency. Yet scientific computing has traditionally favoured higher precision; LineShine’s arrival at the top of the June 2026 TOP500 list with a CPU-based architecture challenges the assumption that high-powered computing must follow the direction set by AI-oriented accelerators, retaining substantial hardware support for double precision \cite{top500June2026}. Despite this, single precision is already used in weather prediction \cite{p20398,eresmaaOperationalForecastProcess2026}, while lower precisions, sometimes combined with stochastic rounding, have produced encouraging results in climate modelling \cite{paxtonClimateModelingLow2022, kimpsonClimateChangeModelling2023}. Mixed-precision methods offer a potential compromise, combining lower-precision computation with higher-precision correction or error assessment to improve performance while retaining the required numerical accuracy \cite{abdelfattahSurveyNumericalLinear2021}. This makes an understanding of the statistical consequences of finite precision all the more necessary.

Many systems for which we rely on simulation-based inference are chaotic, like the atmosphere \cite{lorenzDeterministicNonperiodicFlow1963}, planetary orbits \cite{laskarChaoticMotionSolar1990}, and molecular ensembles \cite{dellagoKolmogorovSinaiEntropyLyapunov1997}. A perturbed trajectory diverges from its original course exponentially fast, so predicting any specific trajectory becomes infeasible over sufficiently long timescales. The workaround, used almost universally, is to stop predicting individual trajectories, and estimate statistical observables of the system. In practice, these quantities are commonly estimated by time averages over an ensemble of simulated initial conditions, with the assumption that sufficient sampling will recover the steady-state statistics of the system. Such approaches and assumptions underpin modern computational methods including ensemble-based molecular dynamics and computational fluid dynamics \cite{wanRapidAccuratePrecise2020, suterRapidAccurateReproducible2025, coveneyMolecularDynamicsProbability2025}.

Birkhoff's theorem gives us a theoretical justification: for an ergodic probability measure $\mu$, time averages of an integrable observable converge to its expectation under $\mu$ for $\mu$-almost every initial state. However, this guarantee concerns the intended continuum dynamics - the set of representable states available on a computer has measure zero, and the resultant simulated trajectories may not be representative. A computer evolves a finite collection of states, with dynamics that can differ substantially from those of the intended continuum system. Reducing precision further coarsens this representation, leaving fewer computer states with which to reproduce the system’s statistical behaviour. Earlier studies established roundoff-induced periodicity and investigated the measures selected by discretised dynamics \cite{beckEffectsPhaseSpace1987, grebogiRoundoffinducedPeriodicityCorrelation1988, goraWhyComputersLebesgue1988, blank1997discreteness}. Diamond et al. provided an early systematic analysis of state-space collapse in discretised chaotic maps \cite{diamondSpatialDiscretizationMappings1993, diamondInvariantMeasureArising1994, diamondBasinAttractionCycles1996, diamondCollapsingEffectsComputation1997}, subsequently developed by Guihéneuf et al. \cite{guiheneufDegreeRecurrenceGeneric, guiheneufCramerDistanceDiscretisations2023a, guiheneufCramerDistanceDiscretizations2024}.

Floating-point arithmetic proves particularly catastrophic in the case of chaotic systems, as the finite-state dynamics may no longer exhibit any of the classical hallmarks of chaos \cite{devaneyIntroductionChaoticDynamical2018}: distinct computed trajectories that pass sufficiently close merge after rounding \cite{guiheneufCramerDistanceDiscretizations2024}, trajectories may fall into disjoint basins of attraction \cite{diamondBasinAttractionCycles1996, boghosianNewPathologySimulation2019}, and the spectrum of unstable periodic orbits is replaced by a finite set of spurious periodic orbits \cite{coveneySharkovskiisTheoremLimits2024}. The resulting concern is whether, at reduced precision, these finite-state pathologies can enter the timescales of practical computation. We develop a finite-state description of the implemented dynamics and derive three propositions connecting their structure to a floor for distributional error, bounds on the duration of accurate averaging, and competing sources of finite-time error. We show that these pathologies are an intrinsic risk of simulating chaotic systems at any fixed precision, and validate our analysis through exhaustive numerical experiments on four chaotic maps with known invariant densities. These examples demonstrate how reducing precision can make statistical errors appear sooner and become more severe, leading to a practical question for reduced-precision simulation: when does further averaging cease to help?

\section*{Floating-point numbers, rounding and the limits of local accuracy}
\label{sec:fpnumbers}

Modern digital computers almost universally represent real numbers using the IEEE 754 floating-point standard \cite{IEEEStandardFloatingPoint}. A normalised binary number takes the form:

\begin{equation}
    x = \pm \left(\sum_{i=0}^{q-1} d_i 2^{-i}\right)2^e, \quad d_i \in \{ 0,1 \}
\end{equation}

where the exponent has a finite range $[e_{\min},e_{\max}]$, and subnormal numbers extend the representation towards zero. Since bit strings are of finite length, the set of floating-point numbers is a finite set of rational numbers. Removing one significand bit approximately halves the number of representable scalar states, so successive bit reductions shrink the state space exponentially.

The result of a floating-point arithmetic operation may not belong to $F(q,e_{\min},e_{\max})$, so an appropriate rounding mode $R: \mathbb{R} \to F(q,e_{\min},e_{\max})$ is used to project back into the set of floating-point numbers. For $x\notin F$, let $x_-<x<x_+$ be its neighbouring
floats and $m=(x_-+x_+)/2$. Deterministic round-to-nearest (RN), the IEEE
754 default, and mode 1 stochastic rounding (SR) \cite{connollyStochasticRoundingIts2021} are defined by
\begin{equation}
R_N(x)=
\begin{cases}
x_-, & x<m,\\
x_+, & x>m,\\
x_{\mathrm{even}}, & x=m,
\end{cases}
\qquad
R_{SR}(x)=
\begin{cases}
x_-, & \text{with probability }\frac{x_+-x}{x_+-x_-},\\[1ex]
x_+, & \text{with probability }\frac{x-x_-}{x_+-x_-}.
\end{cases}
\end{equation}
Here $x_{\mathrm{even}}$ is the neighbour with an even significand; both rules leave exactly representable results unchanged.

Forward error analysis can uniformly bound the discrepancy in a computed update \cite{highamAccuracyStabilityNumerical2002}, and shadowing results can, under suitable hypotheses, guarantee the existence of a close true orbit \cite{hammelNumericalOrbitsChaotic1987, sauerHowLongNumerical1997}. However, these error bounds are pointwise, and cannot in general offer any statistical guarantee on the resulting long-run averages \cite{palmoreShadowingComputableChaotic1987}: the doubling map $2x \text{ mod } 1$ is exactly computable on binary strings, yet every finite bitstring reaches zero \cite{boghosianNewPathologySimulation2019}. The statistics of the implemented dynamics therefore require attention even when the floating-point arithmetic itself introduces no error.

\section*{Functional graphs and Markov chains of computer dynamics}
\label{sec:functionalgraphs}

Let $F=\{x^1,...,x^N\}$ be a finite set of floating-point numbers of size $N$, and $R:\mathbb{R} \to F$ an associated deterministic rounding function. Suppose our simulation takes the form of a simple update rule $x_{n+1}=f(x_n)$, that is, a discrete dynamical system. This encompasses a broad class of algorithms such as fixed-timestep numerical integrators, iterative solvers and recurrent neural networks. $f$ is evaluated on a computer as a composition of $n_f$ elementary floating-point operations, $f_1,...,f_{n_f}$, and rounding is applied after every operation, so the simulated dynamical system is defined as $x_{n+1}= \tilde{f}(x_n) \coloneqq R \circ f_{n_f} \circ \dots \circ R \circ f_1(x_n) \in F$. Whilst we focus on one-dimensional systems in this section for simplicity, in the case of a $d$-dimensional system and $r^{th}$ order recurrence, we may work on the state space $F^{dr}$.

\begin{figure}
    \centering
    \includegraphics[width=0.48\linewidth]{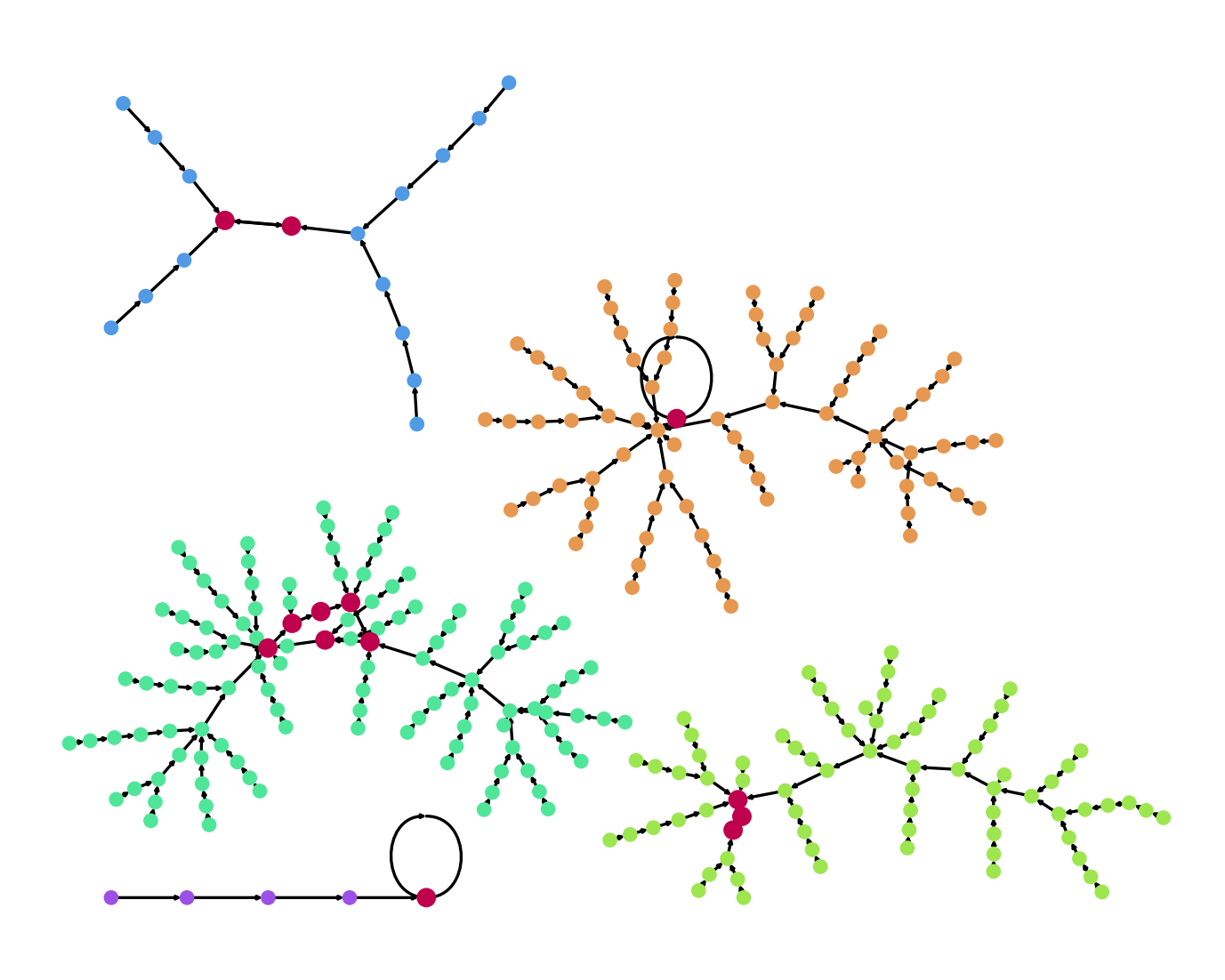}
    \includegraphics[width=0.48\linewidth]{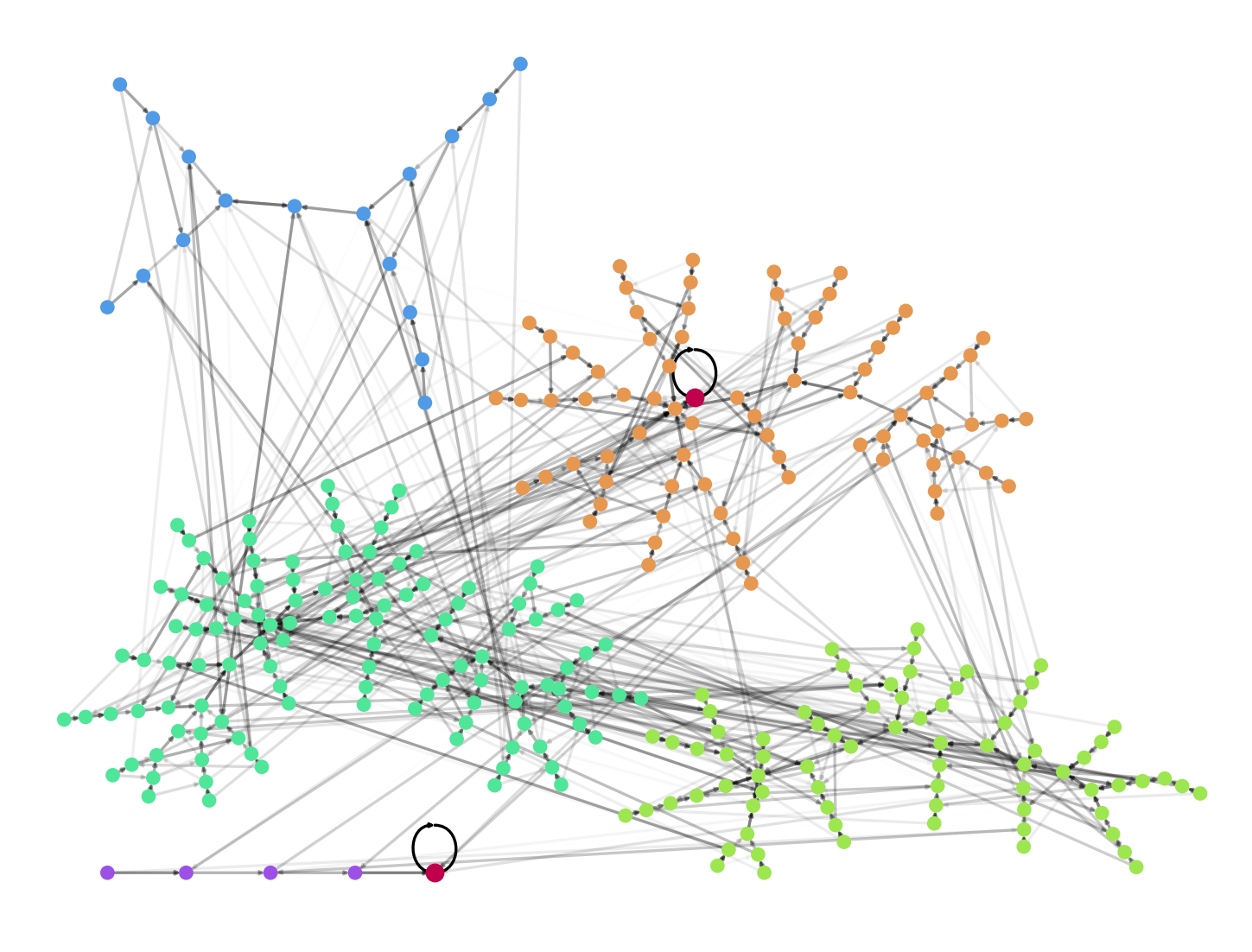}
    \caption{\textbf{Computer dynamics under deterministic and stochastic rounding.} A comparison of the functional graph induced by RN (left), and the Markov chain induced by SR (right) on the logistic map. The invariant sets of both maps are highlighted in red.}
    \label{fig:placeholder}
\end{figure}

One may construct a graph by drawing an outgoing edge from each floating-point number $x^i \in F$ to its computed image under RN, $\tilde{f}(x_i)$; every node has an out-degree of one. In fact, this graph is a \emph{functional graph}, and every connected component consists of trees feeding into exactly one periodic cycle. This structure is naturally encoded in the \emph{transfer operator} $A \in \mathbb{R}^{(N,N)}$ where $A_{ji} = 1$ if $\tilde f(x_i) = x_j$ and $0$ otherwise. This matrix is column-stochastic, and has exactly one 1 in every column, Figure \ref{fig:placeholder} (left) illustrates its trees and periodic cycles.

In the case of stochastic rounding, note that $R_{SR} \circ f_{n_f} \circ \dots \circ R_{SR} \circ f_1(x^i)$ is now a discrete random variable that can take up to $2^{n_f}$ possible values. Repeating the above graph construction induces a finite-state discrete Markov Chain on $F$, with the corresponding column-stochastic transition matrix $A$, where $A_{ji}$ is the probability of moving from $x_i$ to $x_j$. This possesses at least one closed communicating class: a set of mutually accessible states that trap any probability mass entering it. The classes depend on the arithmetic procedure and the floating-point format, Figure \ref{fig:placeholder} (right) illustrates the SR case.

\section*{Distributions on floating-point numbers, and loss of ergodicity}
\label{sec:distributions}

Rather than studying individual trajectories, one may instead ask how probability distributions of states evolve under simulation. This is a natural perspective in statistical mechanics, where the Boltzmann distribution describes ensemble behaviour without reference to any single particle's path. For initial distributions absolutely continuous with respect to the ergodic measure $\mu$, sufficiently long averaging recovers the same invariant statistics; the system should "forget" how the ensemble was initialised.

Consider a discrete probability distribution supported on $F$, represented by probability vector $p$. Its evolution under one step of computer dynamics $\tilde f$ is described by its \emph{pushforward}, $Ap$. Applying $A$ corresponds to pushing forward probability mass at each node according to the edges of the functional graph of the computer dynamics. The long-run behaviour of the dynamics is captured by the \emph{computer invariant measure}, defined as the limiting Cesàro average starting from distribution $p_0$:
\begin{equation}
    \overline{p_\infty}= \lim_{n\to \infty}\frac{1}{n+1}\sum_{i = 0}^n{A^ip_0}
\end{equation}

In the case of RN, we observe that all probability mass will eventually fall into the periodic cycles in their respective connected components in finite time $\tilde n$. This implies that $\overline{p_\infty}$ is a convex combination of stationary distributions associated to each periodic orbit of $A$, which are uniformly weighted on each node of the cycle \cite{boghosianNewPathologySimulation2019}. Under SR, the limiting average combines the stationary laws of closed communicating classes, weighted by probabilities of eventual entry.

This reveals a fundamental departure from the continuous setting: when there are multiple invariant periodic orbits or closed communicating classes, $\overline{p_\infty}$ depends on the choice of initial distribution $p_0$, as probability mass may now fall into different basins of attraction, preventing transitions allowed by the underlying dynamics. The computer invariant measure may not necessarily be unique, imposing an artificial multistability even when the true steady-state is independent of the initial ensemble. Consequently, the choice of initial conditions may introduce a persistent bias in the statistics inferred from simulation, that longer averaging cannot remove.

\section*{Collapse of effective state space}
\label{sec:contraction}

\begin{figure}[t]
\centering

\begin{tikzpicture}[
    every node/.style={inner sep=0pt},
    x=1cm,
    y=1cm
]


\node (h0) at (0,0)
{
    \includegraphics[width=0.26\textwidth]{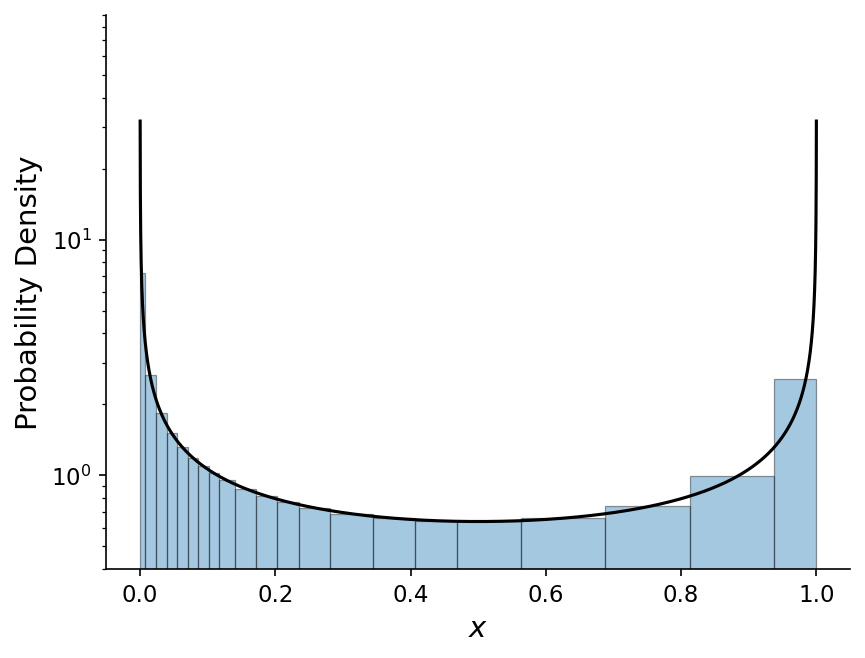}
};

\node (h1) at (5.2,0)
{
    \includegraphics[width=0.26\textwidth]{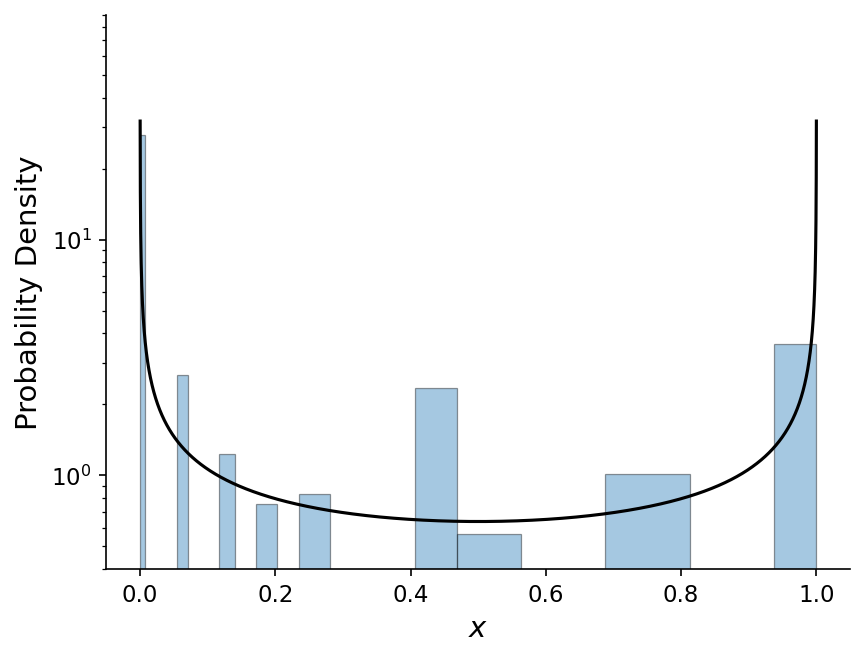}
};

\node at (8.2,0)
{
    {\Huge $\cdots$}
};

\node (h3) at (11.2,0)
{
    \includegraphics[width=0.26\textwidth]{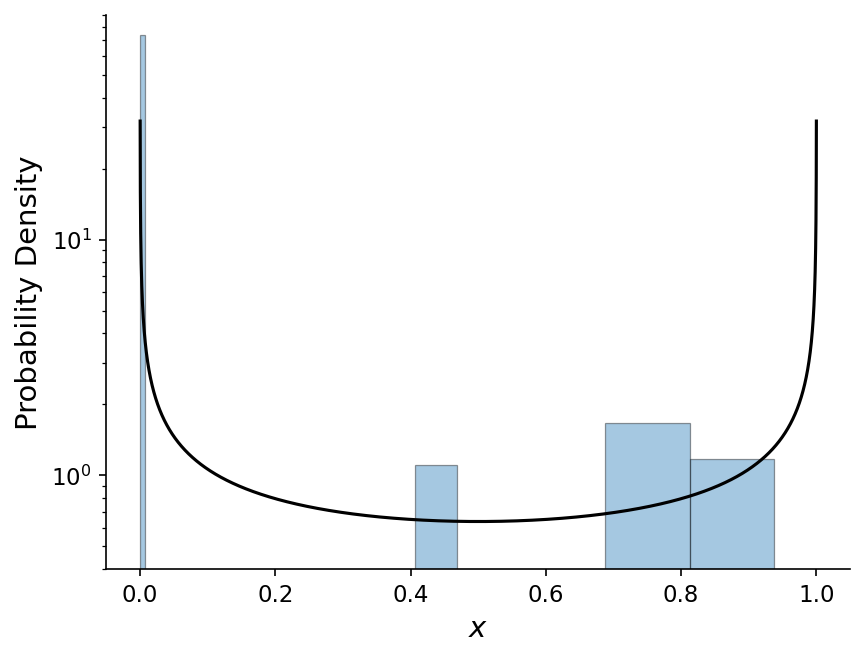}
};


\node (g0) at (0,-4.2)
{
    \includegraphics[width=0.26\textwidth]{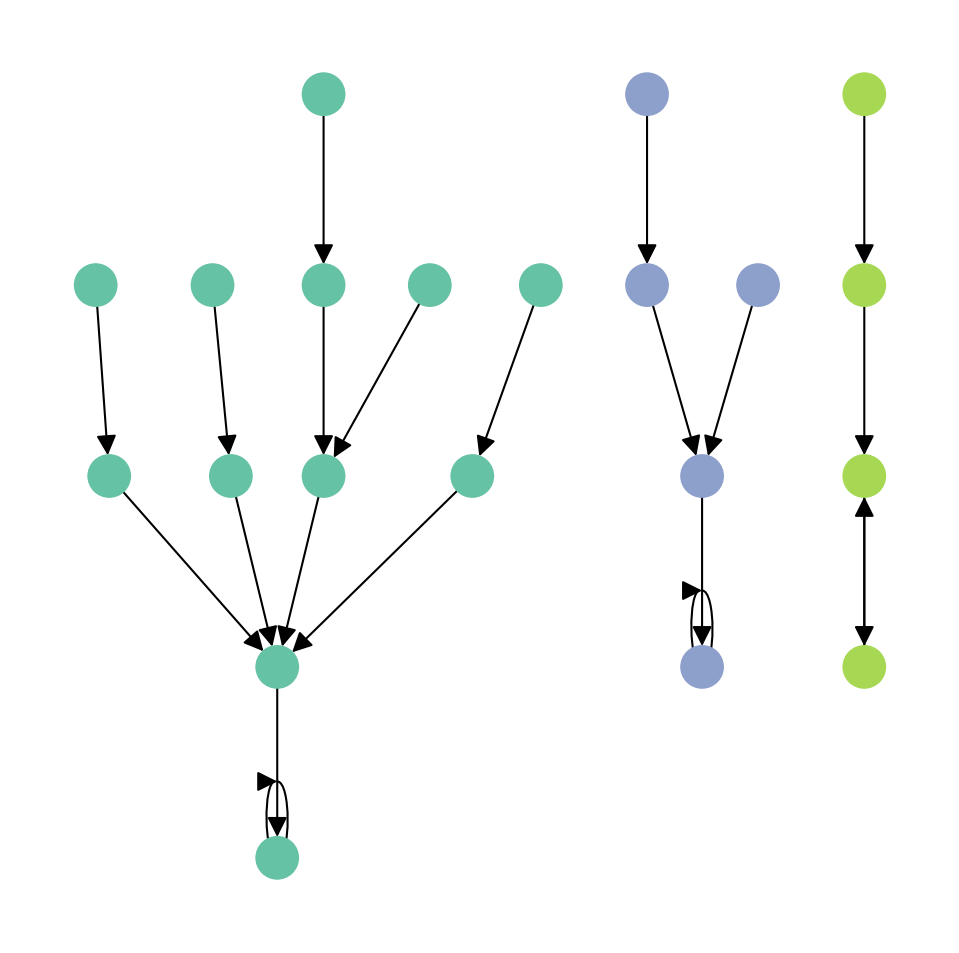}
};

\node (g1) at (5.2,-4.2)
{
    \includegraphics[width=0.26\textwidth]{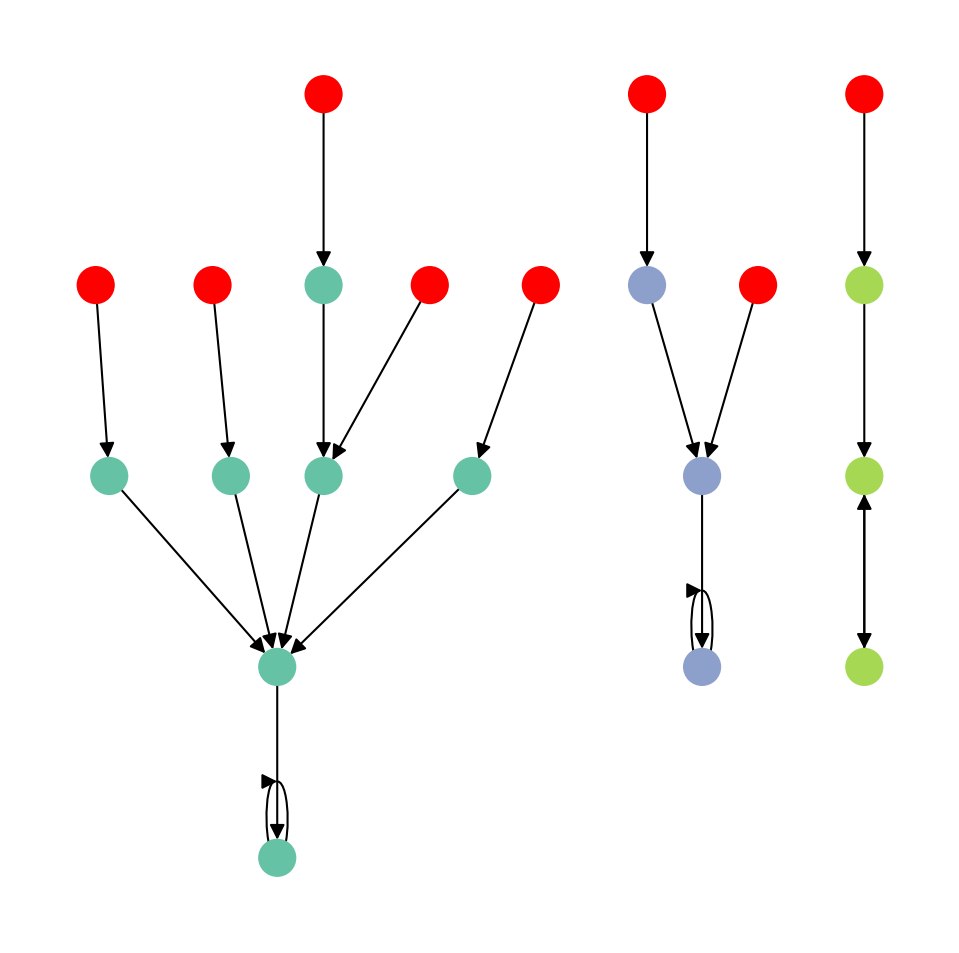}
};

\node at (8.2,-4.2)
{
    {\Huge $\cdots$}
};

\node (g3) at (11.2,-4.2)
{
    \includegraphics[width=0.26\textwidth]{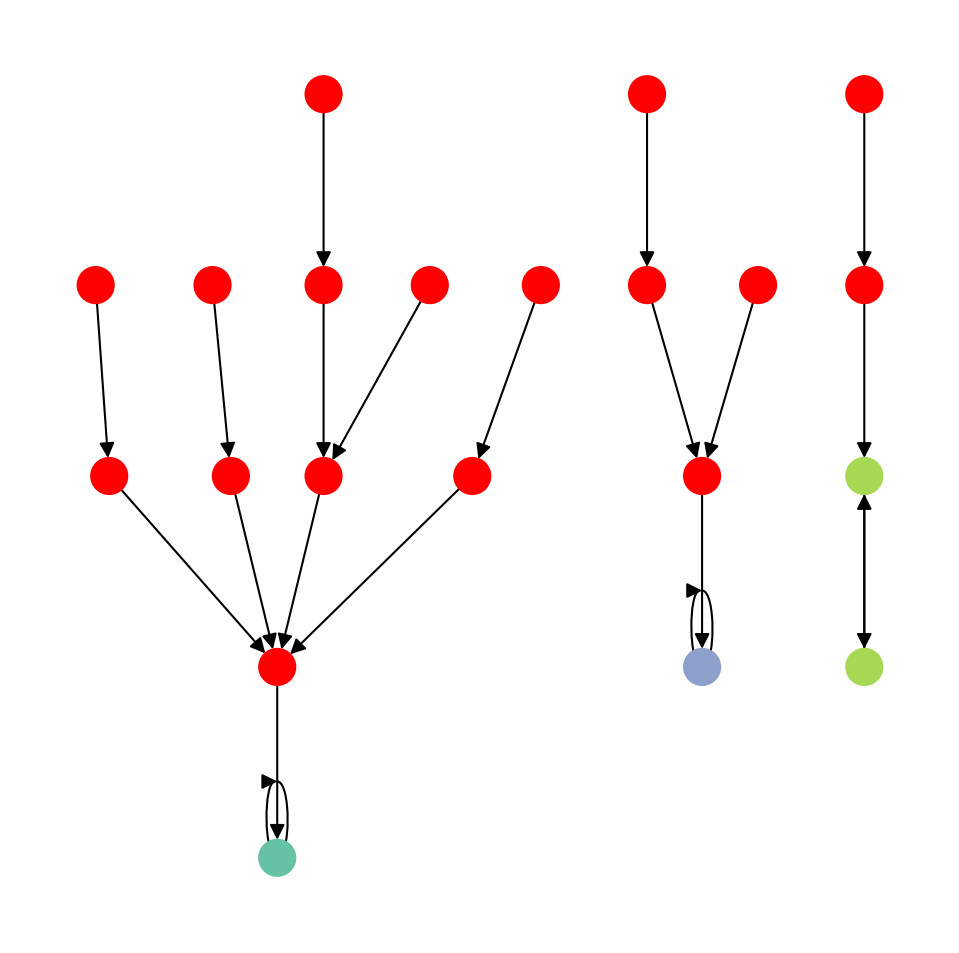}
};


\draw[->, thick]
    ($(h0.east)+(0.2,0)$)
    -- ($(h1.west)+(-0.2,0)$);

\draw[->, thick]
    ($(g0.east)+(0.2,0)$)
    -- ($(g1.west)+(-0.2,0)$);


\node[above=0.2cm of h0]
{
    $p_0$
};

\node[above=0.2cm of h1]
{
    $A p_0$
};

\node[above=0.2cm of h3]
{
    $A^4 p_0$
};

\end{tikzpicture}

\caption{\textbf{Progressive loss of support states.} The logistic density at reduced precision, its pushforward under RN, and at the invariant set after four iterations. Red nodes are unoccupied for all future iterations. Support size monotonically decreases until $\tilde n = 4: K(0) = 21, K(1) = 13, K(4) = 4$.}
\label{fig:transfer_evolution}

\end{figure}

A crucial assumption underpinning numerical simulation is that finite-precision arithmetic preserves the statistical behaviour of the underlying dynamical system. One would hope that the steady state of the true system is preserved under successive iterations of our computer simulation, in other words, that $A^n \tilde{p} \approx \tilde{p}$ for some number of iterations $n \gg 1$. For a non-injective deterministic computer map, this is obstructed via a simple mechanism: the effective state space contracts monotonically until only periodic states remain \cite{diamondCollapsingEffectsComputation1997, yuanCollapsingChaosOne2000, guiheneufCramerDistanceDiscretisations2023a}.

Suppose that $\tilde{p}$ assigns positive probability mass to all states. One iteration of dynamics under RN corresponds to pushing all probability mass forward to their respective successor nodes. Consequently, nodes with zero in-degree will have zero probability mass; since no trajectory can revisit these nodes, they remain unoccupied for all subsequent iterations. Repeating this argument recursively, subsequent iterations continually prune nodes from the trees in the graph until only periodic orbits are left, as illustrated in Figure~\ref{fig:transfer_evolution}. The occupied support therefore has size $K(n) \coloneqq \vert \tilde f^n(F) \vert = \text{rank}(A^n)$, which monotonically decreases until iteration $\tilde{n}$, in which it stabilises to the total number of cycle states, $K(\tilde n)$. As holes accumulate in the support, conservation of probability redistributes mass over the survivors, potentially biasing the expectation of observables.

This phenomenon can also occur under SR. If every intermediate value used to compute a periodic orbit is exactly representable, its evolution remains deterministic under SR. The orbit then forms a closed communicating class: once a trajectory enters it, it cannot leave. These points have the potential to provide an immediate and simple failure mode for stochastic rounding: if such a trapping set is the only closed class reachable from an initial state, the trajectory becomes trapped there with probability one.

The speed of support collapse is entirely dependent on the graph structure, which is determined by the interaction between the arithmetic map and the floating-point format. This graph topology is non-trivial: some chaotic maps induce self-similar trees that collapse to a fixed point exponentially fast \cite{liGraphStructureBakers2025}, while others induce permutations on $F$, and suffer no support collapse at all \cite{liGraphStructureGeneralized2022}. Consequently, the timescale and extent of such finite-precision effects are generally unknown a priori; reducing precision nevertheless constrains the number of states available to sustain transients and periodic orbits.

\section*{Testing the mechanism on four chaotic maps}

We consider binary floating-point formats with $q=1,\dots,24$ significant bits, fixed exponent range and subnormalisation such that the case $q=24$ coincides exactly with IEEE 754 single precision (FP32) \cite{fousseMPFRMultipleprecisionBinary2007}, comprising $1,065,353,217$ states. We also highlight $q=11$ and $q=4$, matching the significand precisions of IEEE binary16 (FP16) and FP8 E4M3 \cite{micikeviciusFP8FormatsDeep2022}. These contain all of the states of their hardware counterparts on this interval, retaining the same resolution above $2^{-14}$ and $2^{-6}$, respectively, while providing additional representable states closer to zero. Under these formats and RN and SR, we construct the pushforward operator for four discrete-time chaotic maps in Table \ref{tab:maps}.

\begin{table}[t]
\centering\small
\begin{tabular}{lll}
\toprule
Map & Update & Invariant density on $[0,1]$\\
\midrule
Logistic & $4x(1-x)$ & $1/[\pi\sqrt{x(1-x)}]$\\[3pt]
Bernoulli & $2x\bmod1$ & $1$\\[3pt]
Generalised Bernoulli & $\varphi x\bmod1$ & $\frac{5+3\sqrt5}{10}\mathbf1_{x<\varphi^{-1}}+\frac{5+\sqrt5}{10}\mathbf1_{x\ge\varphi^{-1}}$\\[4pt]
Generalised Bernoulli & $(4/3)x\bmod1$ & $Z^{-1}\sum_{j\ge0}(4/3)^{-j}\mathbf1_{[0,a_j)}(x)$\\
\bottomrule
\end{tabular}
\caption{Benchmark maps and their absolutely continuous invariant densities. Here $\varphi=(1+\sqrt5)/2$, $a_0=1$, $a_{j+1}=(4a_j/3)\bmod1$, and $Z$ normalises the last density. The non-integer Bernoulli densities follow from the theory of $\beta$-transformations \cite{renyiRepresentationsRealNumbers1957}.}
\label{tab:maps}
\end{table}

Their known absolutely continuous invariant measures (ACIM) provide an analytical reference point, and their one-dimensional state spaces permit computationally feasible exhaustive study. However, these systems should be regarded as a theoretical best-case scenario for inferring statistics from Birkhoff averages: they exhibit exponential mixing towards a unique ACIM. Each system poses distinct challenges for floating-point arithmetic. The logistic map is a canonical discrete chaotic map that exhibits integrable singularities in its invariant density at the interval boundaries \cite{maySimpleMathematicalModels1976}. The Bernoulli map left shifts the binary expansion of a number and hence maps every finite binary string of length $b$ to zero in $b$ iterations \cite{friedmanErgodicPropertiesComputer1988}. The generalised Bernoulli map with parameter $\beta = \varphi \coloneqq \frac{1+\sqrt{5}}{2}$ features a single discontinuity in its invariant density at the irrational point $\varphi^{-1}$ \cite{renyiRepresentationsRealNumbers1957}, while $\beta = \frac{4}{3}$ exhibits countably many discontinuities in its invariant density \cite{hofbauerVShiftsHaveUnique1978}.

\begin{figure}
    \centering
    \includegraphics[width=\linewidth]{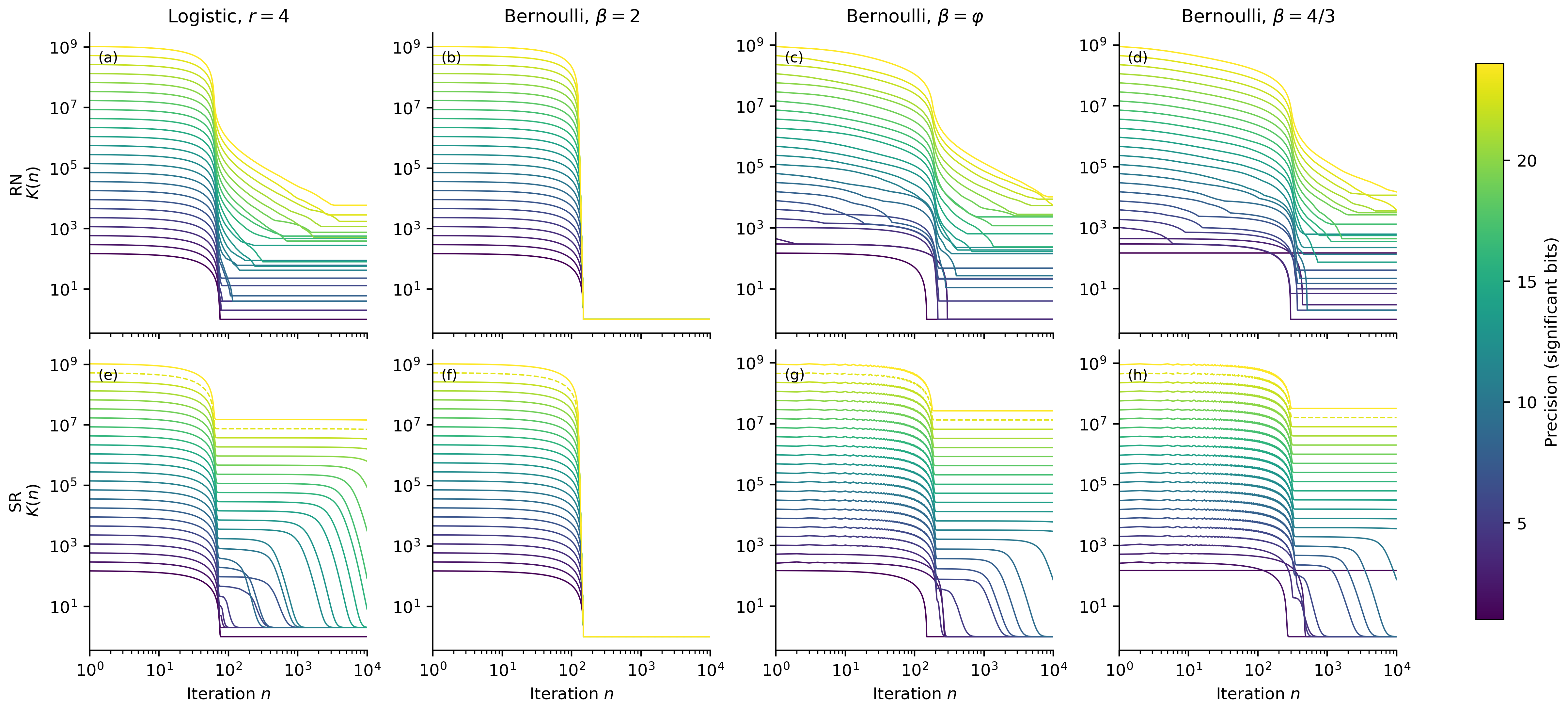}
    \caption{\textbf{Rapid contraction of the effective state space.}
Evolution of $K(n)$, coloured by precision $q$, under RN (top) and SR (bottom).
In both cases, $K(n)/K(0)$ measures probability overlap with the initial equal-state distribution: under RN, $K(n)$ exactly counts reachable states; under SR, it also accounts for states whose occupation probability falls below its initial value.
SR retains greater overlap over extended transients, with plateaus that can precede a further decline as probability concentrates.}
    \label{fig:Kn}
\end{figure}

At each significant-bit precision, $q=1,...,24$, we construct
the RN and SR transfer operators and propagate both the projected invariant law and a projected continuous uniform law for $10^4$ iterations. The first initialisation measures the drift from the desired statistics; the second allows continuum convergence and finite-precision effects to compete. The resulting distributions allow us to compute expected ensemble statistics free from Monte Carlo sampling uncertainty.

We separately propagate the equal-state law $u_j = \frac{1}{N}$, and compute:
\begin{equation}
K(n) = \sum^N_{i=1}{\min{(1, N[A^n u]_i})}
\end{equation}

Each state contributes its probability relative to the initial mass $1/N$, capped at one. In RN, this coincides with the size of the effective state space $\vert \tilde f^n(F) \vert$. Under SR, states with probability below $1/N$ contribute fractionally, distinguishing between reachability and substantial occupation, especially where some nodes may hold negligible probability mass.

Figure \ref{fig:Kn} shows that across all tested maps, the effective state space can contract by several orders of magnitude within a relatively small number of iterations. At $q=24$, logistic RN has fewer than a million reachable states after only 100 updates, removing more than \(99.9\%\) of the initial state space. Reducing precision to $q=11$ leaves $125$ states, while at $q=4$ only $2$ remain. The non-integer Bernoulli maps contract more slowly, both losing over $99\%$ of their initial states within 1000 updates at precisions $q=24,11,4$. Doubling reaches a single state after 149 updates at every tested precision and behaves identically under both rounding rules.

Apart from $\beta=2$, SR sustains considerably greater occupation across the grid over extended transients, with higher precision prolonging an intermediate plateau before a secondary decline. After the same 1000 updates, both non-integer Bernoulli maps retain $K(n)$ between $2\%$ and $3.1\%$ of their initial grid size at $q=24$ and $q=11$, compared with less than $0.1\%$ under RN. At lower precisions, the SR curves approach $K=1$ for the non-integer Bernoulli maps and $K=2$ for the logistic map, consistent with absorption at $0$ and at $\{0,3/4\}$, respectively. In fact, these are the only entirely exact periodic orbits of the corresponding implemented maps, and if they are accessible from every state, then the support almost surely collapses to a fixed size independent of precision.

\section*{A rising floor for statistical error}

In practice, we average values of an observable $\phi$ along a trajectory. After $n$ updates there are $n+1$ total observations; its expected average is the expectation under the Ces\`aro average of the corresponding pushforward measures:
\begin{equation}
\mathbb E\!\left[\frac{1}{n+1}\sum_{i=0}^{n}\varphi(X_i)\right]
=\int\varphi\,d\bar\nu_n,\qquad
\bar\nu_n=\frac{1}{n+1}\sum_{i=0}^{n}\nu_i,\qquad
\nu_i=\sum_{x\in F}[A^ip]_x\,\delta_x.
\end{equation}

We measure statistical discrepancy from the target $\mu$ by the Wasserstein-1 distance $W_1(\mu,\bar\nu_n)$, the minimum transport cost of transporting one distribution to the other with cost of mass times distance. Its relevance to observables follows from the Kantorovich-Rubinstein duality \cite{villaniOptimalTransport2009}:
\begin{equation}
W_1(\mu,\bar{\nu}_n) =
\sup_{\operatorname{Lip}(\varphi)\leq1}
\left|
\int\varphi\,d\bar\nu_n-\int\varphi\,d\mu
\right|
\label{eq}
\end{equation}

Thus $W_1(\mu,\bar{\nu}_n)$ is the largest absolute expected bias among unit-Lipschitz observables, and the bias of an $L$-Lipschitz observable is at most $L \cdot W_1(\mu,\bar{\nu}_n)$; on compact domains such as $[0,1]$, this includes $C^1$ observables such as positive integer moments. This measures the systematic discrepancy remaining after averaging over the initial ensemble, however, a finite sample of trajectories will also have sampling uncertainty.

Precision enters through both the available states and the dynamics on them. Let $S_{\infty}(q)$ be the invariant set of the computer dynamics at precision $q$: under RN, this is the union of periodic states, and under SR, it is the union of closed communicating classes. Define $L_n(q)$ as a lower bound for $W_1(\mu,\bar{\nu}_n)$, i.e.  the smallest worst-case bias in Lipschitz observables obtained by averaging over $n$ iterations. The formal construction and proof is given in the appendix.

\begin{proposition}[Rising error floor for finite-time averages]
\label{prop:wasserstein_floor}
For $\tilde f_q$ under RN or SR on a finite subset of a compact metric space $(X,d)$ and any initial law on $F_q$:
\[
W_1(\mu,\bar{\nu}_n)\geq L_n(q),
\qquad
L_{n+1}(q)\geq L_n(q), \qquad \lim_{n\to\infty}L_n(q)
= L_\infty \coloneqq
\int_X d(x,S_{\infty}(q))\,d\mu(x),
\]
where $d(x,S)$ is the distance to the nearest state in $S$.
\end{proposition}

As the simulation concentrates observations on fewer states, even the most favourable weighting of those observations becomes less able to reproduce the true statistics. Under RN, this restriction arises from accumulating holes in the support; under SR probability drains from transient states to the invariant set. Earlier observations on a larger state space are diluted as the error floor approaches the cost of moving the target distribution to the nearest points in the invariant set.

The initial floor is determined by precision: the lower the precision, the lower the resolution at which we can approximate the steady-state distribution. Projecting the target measure onto $F_q$ assigns the mass of each nearest-neighbour cell $I_i$ to its representative $x_i$:

\begin{equation}
\tilde\mu_q\coloneqq\sum_{x_i\in F_q}\mu(I_i)\delta_{x_i},
\qquad
L_0(q)\coloneqq W_1(\mu,\tilde\mu_q) \leq \max_i \text{diam}(I_i)
\end{equation}

This is the closest measure to $\mu$ supported on the grid; consequently, $L_0$ is an unavoidable quantisation floor, which itself is bounded by the resolution of the grid \cite{singhMinimaxDistributionEstimation2019}, which for our experiments, $\max_i \text{diam}(I_i) = 2^{-q}$ on the interval $[0,1]$ \cite{IEEEStandardFloatingPoint}. Thus the initial error floor becomes exponentially larger at reduced precision, becoming macroscopic at sufficiently low precision.

\begin{figure}
    \centering
    \includegraphics[width=\linewidth]{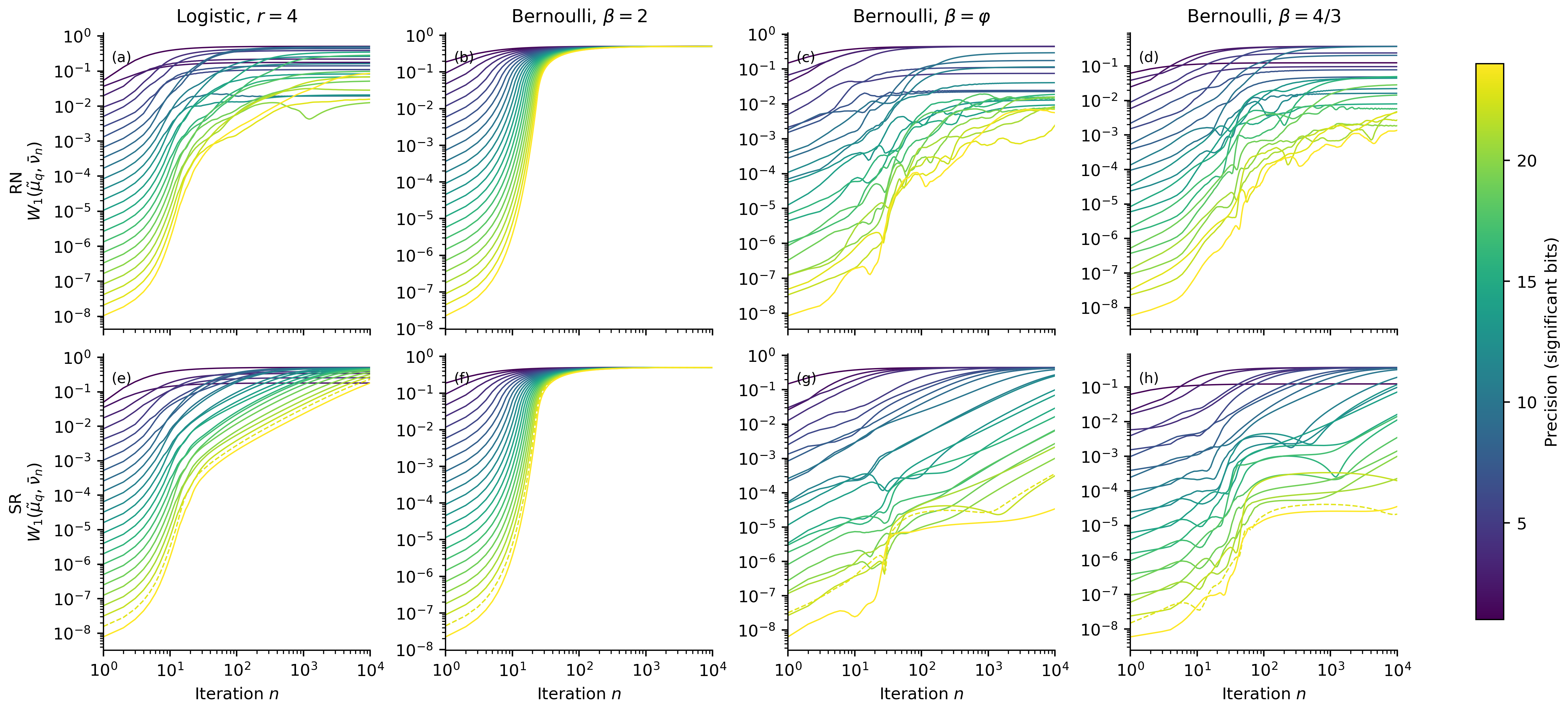}
    \caption{\textbf{Statistical drift from initially accurate statistics.} Starting from the projected invariant distribution, the time-averaged distribution departs from this reference under RN (top) and SR (bottom), with colours indicating precision $q$. Discrepancy is measured by the Wasserstein-1 distance. Lower precision generally brings substantial drift earlier, while the benefit of SR depends on both the map and precision.}
    \label{fig:acim}
\end{figure}

Figure \ref{fig:acim} initialises at $\mu_q$ and plots $W_1(\mu_q,\bar{\nu}_n)$, which differs from the continuum error by at most $2^{-q}$. Departure from this initially accurate distribution exposes the effect of computer dynamics, with lower precision often bringing appreciable drift much earlier. Taking a projected-reference error of $\varepsilon = 0.01$, logistic RN first reaches it after $381$ iterations at $q=24$, $12$ iterations at $q=11$, and only $2$ iterations at $q=4$. Reducing precision thus removes much of the averaging interval available before an already accurate ensemble develops a substantial bias. However, decrease in error is not necessarily monotone with precision: logistic RN has limiting error $0.01636$ at $q=23$ and $0.09686$ at $q=24$.

The non-integer Bernoulli maps show an especially strong precision dependence under SR. At $q=24$, both remain below $0.01$ throughout the $10^4$ snapshots, whereas at $q=11$ the golden-ratio map first crosses it after $214$ iterations and the $4/3$ map after $580$; at $q=4$, just after $1$ and $5$ iterations. Golden-ratio SR's final error rises from $3.39\times10^{-5}$ at $q=24$ to $0.272$ at $q=11$ and $0.447$ at $q=4$. Its advantage over RN reverses: at $q=24$ its final error is about 165 times smaller, but at $q=11$ it is about $2.4$ times larger. The benefit of SR therefore depends strongly on the precision at which it is used.

Precision also affects how long trajectories can retain statistics distinct from their eventual cycles. Let $\tau_q$ be the entry time into the invariant set $S_{\infty}$, and $\overline{\tau}_q=\mathbb{E}_{\nu_0}[\tau_q]$, averaging with respect to the initial distribution $\nu_0$ and transition matrix $A$.

\begin{proposition}[A bound on the duration of accurate averaging]
\label{prop:accurate_averaging_duration}
\begin{equation}
    W_1(\mu,\bar{\nu}_n)
    \geq L_\infty-\frac{D\,\overline{\tau}_q}{n+1}.
\end{equation}
where $D=\operatorname{diam}(F_q)$. In particular, an average with error at most $\varepsilon<L_\infty$ must satisfy
\begin{equation}
    n+1\leq \frac{D\,\overline{\tau}_q}{L_\infty-\varepsilon}.
\end{equation}
\end{proposition}

For a sufficiently small prescribed error tolerance, Proposition 2 places a necessary upper limit on accurate averaging, beyond which transient observations carry too little weight to offset the influence of the rising error floor. Reducing precision can therefore tighten this horizon by shortening transients and raising the limiting error floor, leaving fewer iterations over which representative statistics can be accumulated. 

\section*{Shorter trajectories, better statistics}

\begin{figure}
    \centering
    \includegraphics[width=\linewidth]{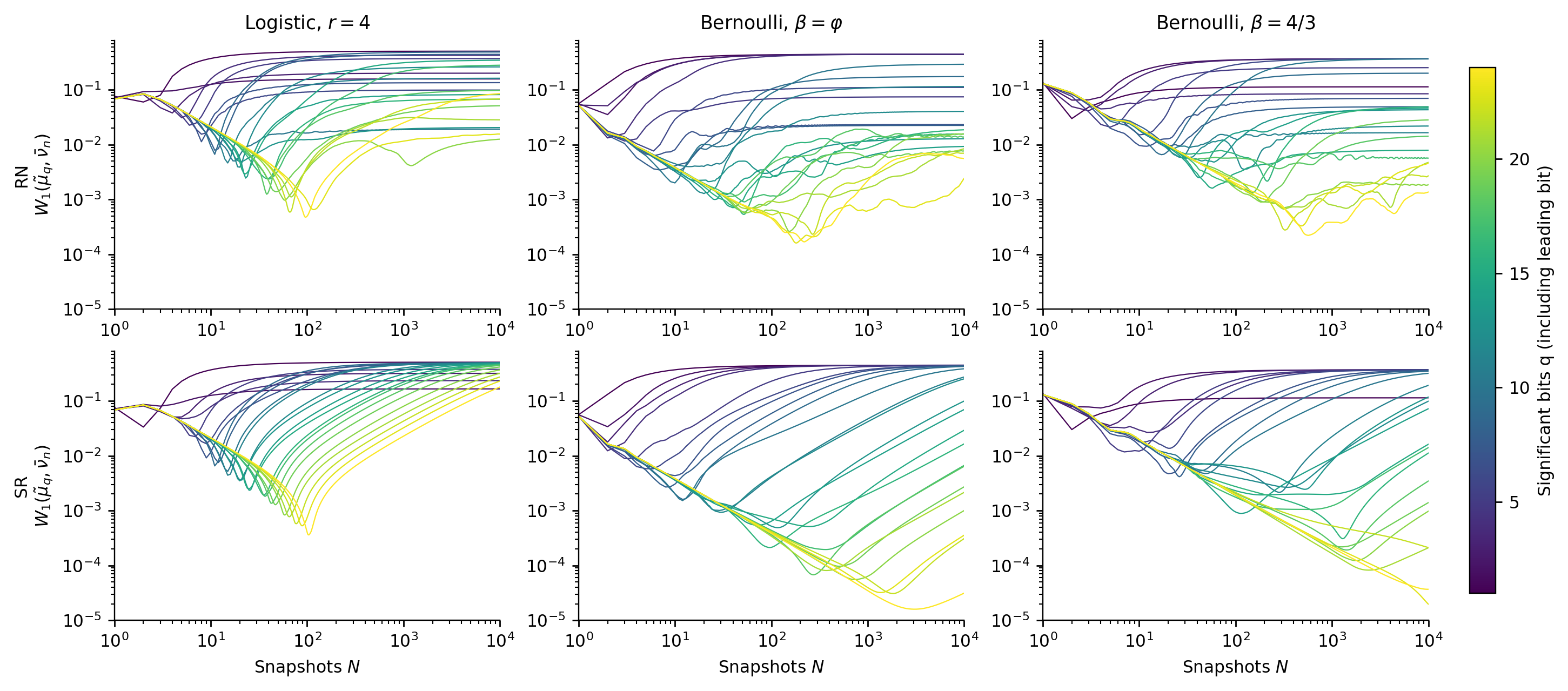}
    \caption{\textbf{Finite-time horizons can beat the "infinite-time" average.} From a projected uniform initial distribution, statistical error can decrease before rising again as finite-precision effects accumulate. The resulting minima depend on precision $q$ (colour) and rounding rule (RN above, SR below), with lower precision often shortening the useful averaging window and limiting the accuracy attained. Error is measured by the Wasserstein-1 distance between the time-averaged distribution and the projected invariant reference. Doubling is omitted because its uniform invariant law gives the same experiment as Fig.~\ref{fig:acim}.}
    \label{fig:uniform}
\end{figure}

In practice, initial conditions are sampled uniformly at random, with the assumption that convergence to steady-state occurs within some fixed number of burn-in iterations. Even if we recover any convergence properties inherited from the true continuum dynamics, we have demonstrated that inevitably, finite-precision effects win out, and the contraction of the effective state space drives the ensemble toward the computer invariant measure.

Figure \ref{fig:uniform} starts from the projected uniform law, allowing relaxation and finite-precision effects to compete. In general, averages appear to improve before deteriorating, so the trajectory length becomes a choice affecting statistical bias. Logistic RN at $q=24$ reaches its best recorded error of $4.78\times10^{-4}$ after $95$ iterations, approximately two hundred times smaller than its limiting error of $0.09687$. At $q=11$ the best error is $4.65\times10^{-3}$ after $14$ iterations, and at $q=4$ it is $3.10\times10^{-2}$ after $4$. These precision reductions make the best estimate progressively worse and bring it earlier; even at $q=4$, the finite average is about $14$ times more accurate than its computer invariant measure counterpart.

This competition can be expressed through a standard Wasserstein distance argument for Lipschitz systems, propagating initial discrepancies and accumulating one-step defects \cite{rudolfPerturbationTheoryMarkov2018}. Let $\rho_i = (f^i)_\# \rho_0$ be the continuum evolution, $W_1(\rho_i, \mu) \leq r_i$ a relaxation bound, and $\delta_0 = W_1(\nu_0, \rho_0)$ the initial representation error.

\begin{proposition}[Three sources of finite-time error]
\label{prop:upper}
Assume $f_\#\mu=\mu$, $f$ is globally $\Lambda$-Lipschitz with $\Lambda\ge1$. Let $\varepsilon_{}=\sup_{x\in F_q} {\sum_{y \in F_q} A_{yx}d(y,f(x))}$ be the largest expected one-step displacement from the continuum map, then:
\begin{equation}
 W_1(\mu,\bar{\nu}_n)\le U_n:=\frac1{n+1}\sum_{i=0}^{n}
 \left[\underbrace{r_i}_{\text{relaxation}}
 +\underbrace{\Lambda^i\delta_0}_{\text{initial representation}}
 +\underbrace{\varepsilon\sum_{k=0}^{i-1}\Lambda^k}_{\text{arithmetic defect}}\right],
 \label{eq:upper}
\end{equation}
\end{proposition}

The bound separates relaxation towards the target statistics from finite-precision effects and accumulation of floating-point arithmetic defects. For convergence of the true dynamics to $\mu$, we must have that $r_i$ decreases to zero. However, the upper bound is eventually driven upwards by both finite-precision and accumulating errors. Although the exponential term may appear conservative, the Bernoulli shift in binary fixed-point arithmetic provides an exact example: instantiated at a uniform distribution, the first and third terms are zero, while the finite-precision error doubles at every iteration until all states collapse to zero.

This suggests that an average taken over a shorter time period may recover statistics closer to the true value, due to sampling from a distribution that has degraded to a lesser extent. If a valid finite-time upper bound satisfies $U_n(q)<L_\infty(q)$, the corresponding average must have smaller distributional error than the infinite-time computer average, placing the global minimum of this error at a finite horizon. Figure~\ref{fig:uniform} exhibits clear finite-time minima whose locations and depths depend on precision, with the comparisons above showing that reducing precision can both shorten the useful averaging window and worsen the best estimate obtained within it. These observations are consistent with earlier findings that longer numerical records need not improve statistical inference \cite{guiheneufCramerDistanceDiscretisations2023a,velichkoWhenMoreData2026}, while relating this behaviour to the redistribution of probability over the computer state space. They also provide a further motivation for ensemble averaging: increasing the number of independent trajectories within a useful finite-time horizon can reduce sampling uncertainty without extending every trajectory into a regime of greater statistical bias. The practical task is therefore to choose trajectory length and ensemble size together, balancing the accuracy of the sampled distribution against the uncertainty and computational cost of estimating its observables.

\section*{Conclusion}
\label{sec:conclusion}

Reducing precision coarsens the representation of the physical state space and reduces the number of states available to transient and periodic trajectories. This restricts the best attainable distributional accuracy before the simulation begins, while subsequent mergers redistribute probability over the surviving states. An initially accurate ensemble can consequently evolve towards considerably wrong statistics before periodicity, and artificial disjoint computational basins can create an initialisation bias absent from the intended equilibrium statistics. 

Our propositions and numerical validation show how this can limit accurate averaging to a finite interval, with errors orders of magnitude below those of the computer average. SR extends some useful intervals substantially while producing greater late-time bias in others. Increasing precision enlarges the representation and changes its cycles, mergers and basins, giving an overall opportunity for improvement but no guarantee of a monotonic reduction in error; neither the onset of this bias nor its growth rate and magnitude is generally known a priori. However, every simulation using deterministic or stochastic rounding on a finite state space is subject to an unavoidable floor on distributional error that cannot decrease. In particular, reducing precision can raise this bound exponentially and shorten the interval available for accurate averaging.

Despite these difficulties, reduced precision has delivered good results across forecasting and atmospheric dynamics \cite{cooperRelationshipNumericalPrecision2020,kimpsonClimateChangeModelling2023,banderierReducedFloatingpointPrecision2024,chenMixedprecisionComputingGRIST2024}. The exponentially larger state spaces of high-dimensional systems may help explain these successes, consistent with the growth of computer periods observed as Lorenz-96 dimension increases \cite{klowerPeriodicOrbitsChaotic2023}. Finite representation and trapping still remain possible, while numerical integration of continuous-time systems introduces a further dependence on timestep: sufficiently small increments can be lost to rounding, causing stagnation \cite{crociEffectsRoundtonearestStochastic2023,paxtonClimateModelingLow2022}. Above this threshold, a smaller stepsize requires more updates over a fixed physical interval, potentially exposing the computation to finite-state effects before enough effectively independent observations are obtained; on the other hand, too large steps can instead cause numerical instability.

Alternative approaches, including Ulam discretisations, Koopman-operator learning and periodic-orbit cycle expansions, offer other ways of recovering statistical properties without relying solely on long trajectory averages. Their reliability requires control of discretisation errors, spectral approximation or orbit-expansion convergence, respectively \cite{bahsounRigorousApproximationDiffusion2016,colbrookAdversarialDynamicalSystems2026,artusoRecyclingStrangeSets1990}. In practice, direct simulation remains the principal route to statistical inference for many large scientific models, allowing selected observables to be estimated without first constructing a sufficiently resolved operator approximation or representative periodic-orbit collection. Although AI-based surrogates are becoming increasingly prominent, those trained on simulated trajectories may inherit and propagate the aforementioned systematic biases in their training distributions.

The hardware context makes these questions increasingly pressing. Accelerator development increasingly prioritises reduced-precision arithmetic, often reducing native FP64 or FP32 throughput \cite{dongarraHardwareTrendsImpacting2024}. NVIDIA’s HGX B200–B300 transition cuts native FP64 throughput by approximately $97\%$ while increasing dense FP4 throughput by $50\%$ \cite{nvidiaHGX2026}. Computational scientists consequently face growing pressure to use these formats for their memory and performance gains; establishing the statistical accuracy delivered by those choices is critical.

Reliable statistical inference therefore requires checking the effects of precision and averaging duration together. A longer trajectory can give a more precise estimate of the wrong distribution, while an ensemble of shorter trajectories may retain more representative statistics. The question for reduced-precision simulation is how long the computation remains statistically useful, and how best to obtain the required observations within that interval.

\section*{Methods}
\label{sec:methods}

The four maps were implemented in C using GNU MPFR \cite{fousseMPFRMultipleprecisionBinary2007} and GMP, with $q=1,\dots,24$ significant binary digits, fixed MPFR exponent settings $\text{emin} = -148, \text{emax} = 128$ and gradual underflow through \texttt{mpfr\_subnormalize} after each rounded operation. Floating-point arithmetic is non-associative; the logistic update evaluates $1-x, 4x$ and their product in that order, and Bernoulli updates multiply by $\beta$, stored in its respective format, then apply modulo $1$.

Every representable state in $[0,1]$, including endpoints, is enumerated by exponent and significand to construct the transfer operator. Under RN, graph traversal identifies all cycles and basins to construct the computer invariant measure, and recursive removal of zero-in-degree states yield the reachable-state counts. Under SR, we enumerate all arithmetic rounding paths and their probabilities using exact rational results and directed MPFR rounding. Notably, SR does not preserve invariant sets, and without correction, may yield unbounded trajectories: final results outside of $[0,1]$ are clamped into the boundaries. All probabilities stored in the transfer operator, their propagation and diagnostics use FP64.

Initial and target distributions are calculated by evaluating closed-form C.D.F. differences on rounding cells for each format, except for $\beta= 4/3$, which truncates the series expansion to $100$ terms; the truncation error is of the order $O(\beta^{-n})$ due to exponential decay of terms \cite{hofbauerVShiftsHaveUnique1978}, which is several orders of magnitude smaller than single-precision machine epsilon.

A sweep over all four maps, 24 precisions, rounding modes and initial distributions was performed on Myriad. At each step, the instantaneous and Cesàro averaged Wasserstein-1 distances are computed by integrating the absolute difference of the cumulative distributions over the floating-point grid. Computations use sparse incoming-edge propagation with OpenMP, and the $\beta=2$ case provides an exact-arithmetic RN–SR control.

\section*{Appendices}

\appendix
\section*{Proof of Proposition 1}
Fix the precision and suppress the subscript $q$. Let $(X,d)$ be compact, let $F\subset X$ be nonempty and finite, and let $A$ be the column-stochastic transition matrix of the computer dynamics. 

Let $S_\infty$ be the union of all closed communicating classes of $A$, and let $S_i$ be the states reachable from $F$ in exactly $i$ updates along positive-probability transitions. These sets satisfy:
\[
S_\infty\subseteq S_{i+1}\subseteq S_i,
\qquad S_0=F.
\]
In the non-deterministic case, $\bigcap_i S_i$ may contain transient states; define the worst-case survival probability $s_i$:
\[
s_i:=\max_{x\in F}\mathbb P_x(\tau>i),
\qquad
\tau:=\inf\{i\ge0:X_i\in S_\infty\}
\]
where $\mathbb P_x$ denotes the law of the chain started at $x$. Since every state of a finite chain can reach a closed class, there exists $m\ge1$ and $a>0$ such that the probability of entering $S_\infty$ within $m$ updates is at least $a$, from every transient state. The Markov property gives $s_{km}\le(1-a)^k$, so $s_i\downarrow0$.

Define the admissible laws at update $i$ by
\[
\mathcal D_i
:=
\left\{
\eta\in\mathcal P(S_i):
\eta(F\setminus S_\infty)\le s_i
\right\}.
\]
These sets are compact, convex and nested, and contain every evolved law $\nu_i=A^i\nu_0$. We note that in the deterministic case, $s_i= \mathbbm{1}_{i<\tilde n}$, therefore $\mathcal D_i = \mathcal P(S_i)$.

With $\text{Prob}(S)$ denoting probability measures on $S$, define:
\begin{equation}
 \mathcal C_n=\left\{\frac1{n+1}\sum_{i=0}^{n}\eta_i:
              \eta_i\in\mathcal D_i\right\},\qquad
 L_n=\min_{\eta\in\mathcal C_n}W_1(\mu,\eta).
\end{equation}
The minimum exists by compactness, and
$\overline{\nu}_n\in\mathcal C_n$ implies
$W_1(\mu,\overline{\nu}_n)\ge L_n$.

To show monotonicity, take $\eta_i\in\mathcal D_i$ for
$i=0,\ldots,n+1$ and define
\[
\xi_i
=
\frac{n+1}{n+2}\eta_i+\frac{1}{n+2}\eta_{n+1},
\qquad 0\le i\le n.
\]
Nesting and convexity give $\xi_i\in\mathcal D_i$, while
\[
\frac{1}{n+1}\sum_{i=0}^n\xi_i
=
\frac{1}{n+2}\sum_{i=0}^{n+1}\eta_i.
\]
Thus $\mathcal C_{n+1}\subseteq\mathcal C_n$, and consequently
$L_{n+1}\ge L_n$. Finally, write:
\[
Q:=\int_X d(x,S_\infty)\,\mathrm d\mu(x),
\qquad D:=\operatorname{diam}(X).
\]
The nearest-point projection of $\mu$ onto $S_\infty$ belongs to every $\mathcal C_n$ and has transport cost $Q$, so $L_n\le Q$.
Conversely, every $\eta\in\mathcal C_n$ satisfies
\[
\eta(F\setminus S_\infty)
\le\frac{1}{n+1}\sum_{i=0}^n s_i.
\]
Testing Kantorovich--Rubinstein duality with the $1$-Lipschitz
function $d(\,\cdot\,,S_\infty)$ therefore gives
\[
W_1(\mu,\eta)
\ge Q-\int_X d(x,S_\infty)\,\mathrm d\eta(x)
\ge Q-\frac{D}{n+1}\sum_{i=0}^n s_i.
\]
Hence
\[
Q-\frac{D}{n+1}\sum_{i=0}^n s_i
\le L_n\le Q.
\]
Since $s_i\to0$, its Cesàro average tends to zero, proving
$L_n\to Q=L_\infty$.

\section*{Proof of Proposition 2}

Recall from Proposition~1 that for $\overline{\nu}_n \in\mathcal C_n$:
\[
W_1(\mu,\overline{\nu}_n)
\ge
L_\infty-
\int_X d(x,S_\infty) \,\mathrm d\overline{\nu}_n.
\]

We now derive an alternative bound for the second integral. Define the expected first-entry time into $S_\infty$:
\[
\tau:=\inf\{i\ge0:X_i\in S_\infty\},
\qquad
\overline{\tau}:=\mathbb E_{\nu_0}[\tau].
\]
Here $X_0\sim\nu_0$, and the expectation is also taken with respect to the transfer operator $A$. The expected entry time is finite by the finite-chain argument in the proof of Proposition~1. Since $S_\infty$ is closed, a trajectory cannot leave it after entry. Thus, among the $n+1$ observations $X_0,\ldots,X_n$, exactly
$\min\{\tau,n+1\}$ lie outside $S_\infty$:
\[
\sum_{i=0}^n
\mathbf 1_{\{X_i\notin S_\infty\}}
=
\min\{\tau,n+1\}.
\]
The Cesàro law
$\overline{\nu}_n=(n+1)^{-1}\sum_{i=0}^n\operatorname{Law}(X_i)$
is the distribution obtained by selecting an observation time
uniformly from $\{0,\ldots,n\}$, independently of the trajectory.
Its mass outside $S_\infty$ is therefore the expected fraction
of observations made before entry:
\[
\overline{\nu}_n(F\setminus S_\infty)
=
\frac{\mathbb E_{\nu_0}[\min\{\tau,n+1\}]}{n+1}
\le
\frac{\overline{\tau}}{n+1}.
\]

Writing $D=\operatorname{diam}(X)$, the function $\varphi$ vanishes
on $S_\infty$ and is bounded above by $D$ elsewhere. Hence
\[
\int_X\varphi\,\mathrm d\overline{\nu}_n
\le
D\,\overline{\nu}_n(F\setminus S_\infty)
=
\frac{D\,\mathbb E_{\nu_0}[\min\{\tau,n+1\}]}{n+1}.
\]
Combining these estimates gives
\[
W_1(\mu,\overline{\nu}_n)
\ge
L_\infty-
\frac{D\,\mathbb E_{\nu_0}[\min\{\tau,n+1\}]}{n+1}
\ge
L_\infty-\frac{D\,\overline{\tau}}{n+1}.
\]
In particular, if
$W_1(\mu,\overline{\nu}_n)\le\varepsilon<L_\infty$, then
\[
n+1\le
\frac{D\,\overline{\tau}}{L_\infty-\varepsilon}.
\]

\section*{Proof of Proposition 3}

Let $P=f_\#$ denote the pushforward under the true continuum map, and let $A$ be the transfer operator of the computer dynamics. Write
\[
\rho_i=P^i\rho_0,
\qquad
\nu_i=A^i\nu_0,
\qquad
e_i:=W_1(\nu_i,\rho_i),
\]
so that $e_0=\delta_0$. Recall that
$W_1(\rho_i,\mu)\le r_i$.

We first compare one computer update with one continuum update
from the same input law. For any $\nu\in\mathcal P(F)$, sample
$x\sim\nu$ and then a computer successor $Y$ with conditional
probabilities $\mathbb P(Y=y\mid x)=A_{yx}$. The pair $(Y,f(x))$
is a coupling of $A\nu$ and $P\nu$, giving
\[
W_1(A\nu,P\nu)
\le
\sum_{x\in F}\nu(\{x\})
\sum_{y\in F}A_{yx}\,d(y,f(x))
\le\varepsilon,
\]
where
\[
\varepsilon:=
\max_{x\in F}\sum_{y\in F}A_{yx}\,d(y,f(x))
\]
is the largest expected one-step discrepancy.

Next, the $\Lambda$-Lipschitz property of $f$ implies:
\[
W_1(P\alpha,P\beta)\le\Lambda W_1(\alpha,\beta).
\]
Indeed, pushing any coupling of $\alpha$ and $\beta$ through
$(f,f)$ multiplies its transport cost by at most $\Lambda$.
The triangle inequality therefore gives
\[
\begin{aligned}
e_{i+1}
&=W_1(A\nu_i,P\rho_i)\\
&\le W_1(A\nu_i,P\nu_i)
   +W_1(P\nu_i,P\rho_i)\\
&\le\varepsilon+\Lambda e_i.
\end{aligned}
\]
Iterating this recurrence from $e_0=\delta_0$ yields
\[
e_i\le
\Lambda^i\delta_0
+\varepsilon\sum_{k=0}^{i-1}\Lambda^k,
\]
where the sum is zero when $i=0$. Thus the discrepancy at update
$i$ consists of the amplified initial representation error and
the accumulated one-step discrepancies.

Comparing with the target measure now gives
\[
\begin{aligned}
W_1(\mu,\nu_i)
&\le W_1(\mu,\rho_i)+W_1(\rho_i,\nu_i)\\
&\le r_i+\Lambda^i\delta_0
+\varepsilon\sum_{k=0}^{i-1}\Lambda^k.
\end{aligned}
\]
Finally, convexity of Wasserstein distance under averaging implies
\[
W_1\!\left(\mu,\frac{1}{n+1}\sum_{i=0}^n\nu_i\right)
\le
\frac{1}{n+1}\sum_{i=0}^n W_1(\mu,\nu_i).
\]
This follows by averaging couplings of $\mu$ with each $\nu_i$:
the resulting coupling has marginals $\mu$ and
$\overline{\nu}_n$. Consequently,
\[
W_1(\mu,\overline{\nu}_n)
\le
\frac{1}{n+1}\sum_{i=0}^n
\left[
r_i+\Lambda^i\delta_0
+\varepsilon\sum_{k=0}^{i-1}\Lambda^k
\right]
=:U_n,
\]
as required.

\bibliographystyle{unsrt}  

\bibliography{bibliography}

\section{Acknowledgements}
This work was supported by the UK Engineering and Physical Sciences Research Council (EPSRC) grant EP/Y034767/1. We thank the UCL Myriad High Performance Computing Facility for access to its resources. We also thank Jeroen Lamb and Chengqing Li for helpful discussions, and the organisers of the International Conference on Dynamical Systems at IMPA and Operator Methods for Dynamical Systems at the Isaac Newton Institute for opportunities to present this work.

\end{document}